\documentclass[prl,twocolumn]{revtex4-1}
\usepackage{graphicx}
\usepackage{color}
\usepackage{amssymb, amsmath, amsbsy}
\usepackage[applemac]{inputenc}

\begin{document}
\title{Non-Abelian gauge potentials in graphene bilayers}
\author{P. San-Jose$^{1}$, J. Gonz\'alez$^{1}$ and F. Guinea$^{2}$  \\}
\affiliation{$^1$Instituto de Estructura de la Materia (IEM-CSIC), Serrano 123,
        28006 Madrid, Spain\\
        $^2$Instituto de Ciencia de Materiales de Madrid (ICMM-CSIC), Cantoblanco, 28049 Madrid, Spain
        }

\date{\today}


\begin{abstract}
We study the effect of spatial modulations in the interlayer hopping
of graphene bilayers, such as those that arise upon shearing or twisting. We show that their single-particle physics, characterized by charge accumulation and recurrent formation of zero-energy bands as the pattern period $L$ increases, is governed by a non-Abelian gauge potential arising in the low-energy electronic theory due to the coupling between layers. We show that such gauge-type couplings give rise to a potential that, for certain discrete values of $L$,  spatially confines states at zero energy in particular regions of the Moir\'e patterns. We also draw the connection between the recurrence of the flat zero-energy bands and the non-Abelian character of the potential.
\end{abstract}


\maketitle

{\em Introduction.---}
The discovery of graphene, the material made of a one-atom-thick
carbon layer, has provided the realization of a system where the electrons
have conical valence and conduction bands, therefore behaving as massless
Dirac fermions \cite{Novoselov:N05,Zhang:N05,Neto:RMP09}.
A remarkable feature of graphene is
that deformations of its honeycomb lattice may produce a similar effect to
that of gauge potentials in the low-energy Dirac theory \cite{Gonzalez:NPB93}.
Recently, it has been shown that the local in-plane deformations induced by strain
can be mimicked by an effective vector potential, which may give rise to the analogue
of Landau levels in the deformed graphene
sheet \cite{Guinea:NP09}.

In this paper we show that the effect of modulations in the interlayer hopping
of graphene bilayers can be represented in general by a non-Abelian background gauge potential
in the low-energy electronic theory, and that it is responsible for the zero energy charge density waves and the dispersionless minibands,  predicted by theory, 
and recently measured \cite{Luican:PRL11}. The vector components of the potential
take values in the space of SU(2) matrices, which correspond to rotations
in the Hilbert space of the two layers. This kind of non-Abelian gauge
fields \cite{Wilczek:PRL84} is relatively rare in a condensed-matter context \cite{Moore:NPB91, Kitaev:AOP03,San-Jose:PRB08}, but it is quite
relevant in subatomic physics, being responsible for the interaction
between matter fields. The proton and the neutron, for instance, compose
an isospin SU(2) doublet. It was proposed long ago that an ideal experiment
of scattering of these particles onto a non-Abelian flux line should lead to the transfer
of protons into neutrons and vice versa \cite{Wu:PRD75}. In general, matter
fields pick up a matrix-valued `phase' in their propagation in a non-Abelian gauge field. Interference of such matrix-valued phase along two indistinguishable paths (as opposed to the conventional U(1) phase) leads to an intriguing non-Abelian generalization of the Aharonov-Bohm effect. In our context, this would manifest as coherent layer polarization induced by the interference of two SU(2) phases acquired along the two paths.


Experimental realizations of non-Abelian gauge potentials have been proposed before
in the study of ultracold atoms \cite{Osterloh:PRL95,Ruseckas:PRL95}.
Investigations have addressed in particular the influence of the non-Abelian gauge
potentials in the development of the Landau levels produced by a conventional magnetic
field \cite{Goldman:PRA09,Estienne:NJOP11}.
However, the question of
whether pure non-Abelian gauge fields may lead to a phenomenology similar to the magnetic
localization of Landau states remains open.
Our investigation sheds light
on this question, showing that it is possible to develop a zero-energy level of
spatially confined states as a consequence of the non-Abelian gauge potential, provided that the fermion fields
return to the original internal state around a closed path.


The effective non-Abelian gauge potentials that arise
in the bilayers have actually a genuine applied interest, since they induce periodic spatial
confinement of electronic states. Indeed, we will see that the one-dimensional (1D)
modulation of the interlayer tunneling leads to the confinement of electronic
states into narrow 1D channels. We will also extend our approach to the description of
twisted bilayers \cite{LopesdosSantos:PRL07, Morell:PRB10,Trambly-de-Laissardiere:NL10,
Mele:PRB10,Bistritzer:PNAS11,Gail:PRB11,Kindermann:11, Mele:11, Morell:11},
where the non-Abelian gauge potential turns out to confine low energy electrons
into a triangular array of quantum dots. The problem of confinement of electronic states
has particular relevance, given that scalar potential barriers are not effective
to constrain the propagation of the electrons in graphene \cite{Katsnelson:NP06}, which makes
the non-Abelian gauge potentials proposed in this paper an interesting alternative to
the confinement and manipulation of electronic states in graphene devices.

\begin{figure}
\includegraphics[height=4.4cm]{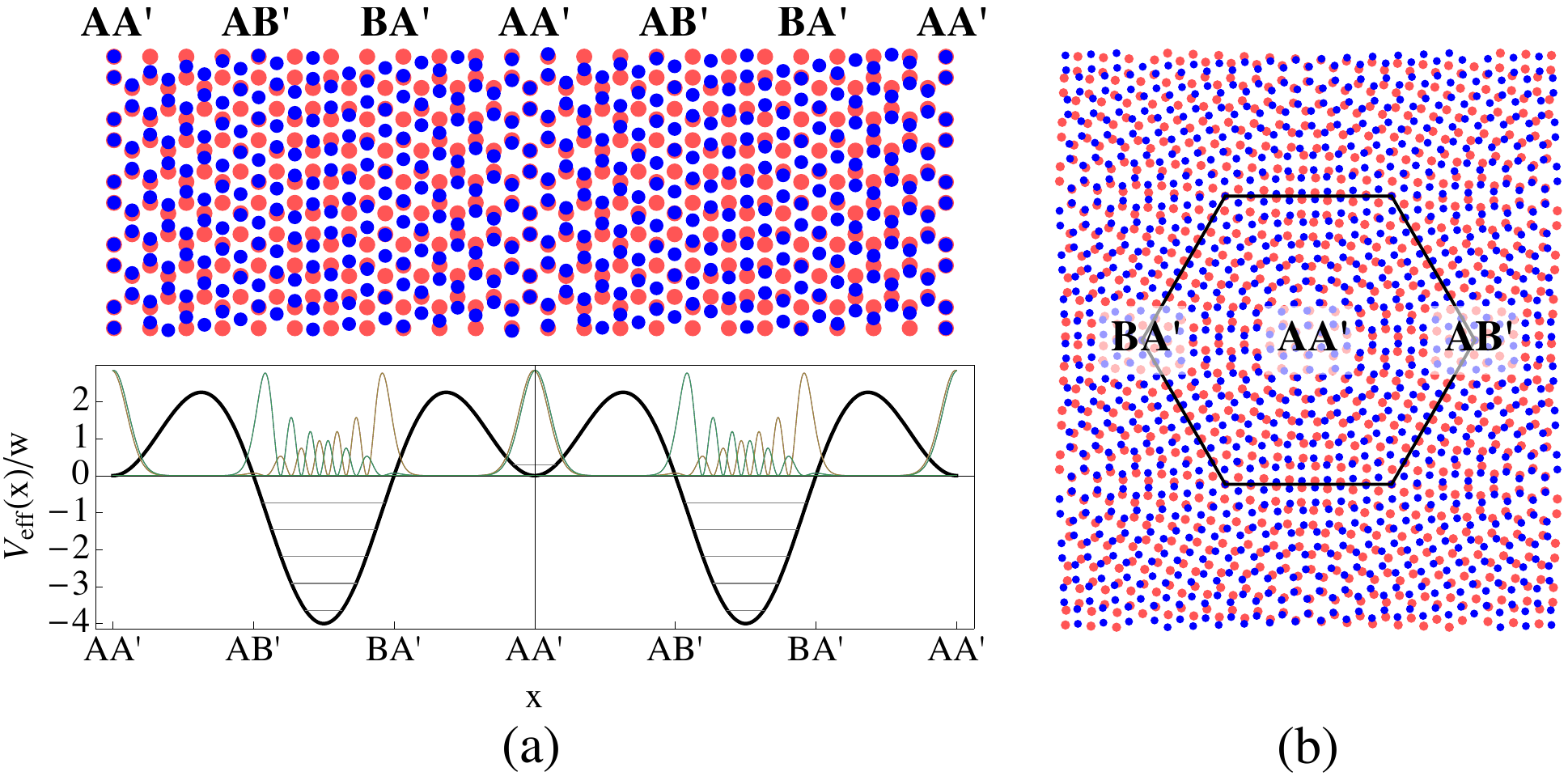}
\caption{Moir\'e patterns of (a), top, sheared bilayer (showing the alternation between
$AA'$, $AB'$ and $BA'$ stackings, and (b) twisted bilayer, where the hexagonal supercell
and the different types of stacking have been marked. (a), bottom, shows the effective
potential $V_\mathrm{eff}(x)$ arising from the non-Abelian gauge potential
$\hat{\mathbf{A}}$, together with a typical zero-energy state confined between the
$AB'$ and $BA'$ regions, and a finite energy state concentrated around $AA'$.}
\label{one}
\end{figure}

{\em Model.---} The simplest realizations of a non-Abelian gauge potential
are found by means of a modulated mismatch in the relative position of the two
lattices of a bilayer, obtained either by applying strain or shear in one of the
layers or by relative rotation between the two layers. In both instances,
the resulting mismatch produces characteristic Moir\'e patterns, see Fig. \ref{one},
which reflect the spatial alternation between $AA'$-type stacking (perfect
alignment of the atoms in the two layers) and $AB'$-type, $BA'$-type (Bernal)
stacking, where $A(')$ and $B(')$ correspond to the two sublattices of the lower
(upper) lattice.

At energies $\varepsilon \lesssim 1$ eV, the Moir\'e electron system is described
by Dirac  fermions on each layer, coupled by a position dependent interlayer hopping
amplitude. The Hamiltonian takes the form \cite{LopesdosSantos:PRL07,Santos:12}\footnote{The absence of a mass term in the diagonal is justified by the symmetry between the two sub-lattices within each layer, which is preserved for smooth Moiré patterns.}
\begin{eqnarray}
H=
v_F\left(\begin{array}{cccc}
0 & \Pi_+^\dagger & V_{AA'}(\mathbf{r}) & V_{AB'}(\mathbf{r}) \\
 \Pi_+ & 0 & V_{BA'}(\mathbf{r}) & V_{AA'}(\mathbf{r}) \\
 V_{AA'}^\star(\mathbf{r}) & V_{BA'}^\star(\mathbf{r}) & 0 &\Pi_-^\dagger \\
 V_{AB'}^\star(\mathbf{r}) & V_{AA'}^\star(\mathbf{r}) &  \Pi_- & 0
 \end{array}\right)
\label{H}
\end{eqnarray}
where $\Pi_\pm \equiv-i\partial_x+\partial_y \mp ( \tilde{A}_x + i \tilde{A}_y )$.
The spatially modulated interlayer coupling functions $V$ arise from the Moir\'e pattern
formation, and the intra-layer \emph{Abelian} gauge field $\pm \mathbf{\tilde{A}}$
describes the strains in each layer. These strains lead to constant gauge fields in our case,
$\mathbf{\tilde{A}} = \Delta \mathbf{K}/2$. Note that, as discussed later, this model also
describes twisted bilayers, in which the interlayer Dirac cone shift $\Delta \mathbf{K}$
arises due to the relative twist between layers, not strains.
Since $\mathbf{\tilde{A}}$ is anyhow uniform, we can gauge it away by a transformation
$U = \exp ( (i/2) \tau_3 \Delta \mathbf{K}\cdot\mathbf{r} )$,
where $\tau_3$ is a Pauli matrix which operates on the layer index.
This transforms consequently the interlayer couplings into
$\widetilde{V}_{ij} ( \mathbf{r} )= V_{ij} ( \mathbf{r} ) e^{-i \Delta \mathbf{K}\cdot \mathbf{r}}$.


{\em Low-energy theory of sheared bilayers.---} We consider first
the instance in which shear $u_{xy}$ is applied along the
$AB$ bonds of a given layer section ($y$ direction).
Then a 1D Moir\'e pattern is produced in the orthogonal $x$ direction, smoothly
alternating between $AA'$, $BA'$ and $AB'$ stacking as shown in Fig. \ref{one}(a).
The corresponding hopping amplitudes are
related by $\widetilde{V}_{AA'}(x)=\widetilde{V}_{BA'}(x-L/3)=\widetilde{V}_{AB'}(x+L/3)$, where
$\widetilde{V}_{AA'}(x)\approx (w/v_F) \left[1+2\cos(2\pi x/L)\right]$ using a single-harmonic
approximation \cite{LopesdosSantos:PRL07} (the interlayer coupling is $w \approx t_\perp / 3 = 0.11$ eV,
where $t_\perp$ is the interlayer hopping).

To assist in interpreting the role of the different interlayer couplings, we define the functions
$A_x (x) = -(\widetilde{V}_{AB'}(x)+\widetilde{V}_{BA'}(x))/2 $ and
$A_y (x) = (\widetilde{V}_{AB'}(x)-\widetilde{V}_{BA'}(x))/2 $. Then $\widetilde{V}_{AB'} = -A_x + A_y$,
$\widetilde{V}_{BA'} = - A_x - A_y$, and it becomes clear that $A_x, A_y$ act as
off-diagonal vector potentials. Taking Pauli matrices $\boldsymbol{\sigma}$ in the $AB$
pseudospin space and $\boldsymbol{\tau}$ in the space of the two layers, we
may recast Eq. (\ref{H}) into
\begin{equation}\label{HA}
H=v_F \boldsymbol{\sigma}\cdot(-i\boldsymbol{\partial}-\hat{\mathbf{A}})+v_F \widetilde{V}_{AA'}\tau_1
\end{equation}
where we have introduced the gauge potential
$\hat{\mathbf{A}}=(A_x \tau_1 ,A_y \tau_2)$, which induces a precession of the layer index as an electron moves in real space.
This $\hat{\mathbf{A}}$ is \emph{non-Abelian}, since $[\hat{\mathbf{A}}(\mathbf{r}),\hat{\mathbf{A}}(\mathbf{r}')]\neq 0$ in general (see also Ref. \onlinecite{Son:PRB11}). This formulation highlights the
different nature of the $\widetilde{V}_{AA'}$ coupling, that acts rather like a scalar
potential (proportional to the unit matrix $\sigma_0$).

\begin{figure}[t]
\begin{center}
\includegraphics[height=4.7cm]{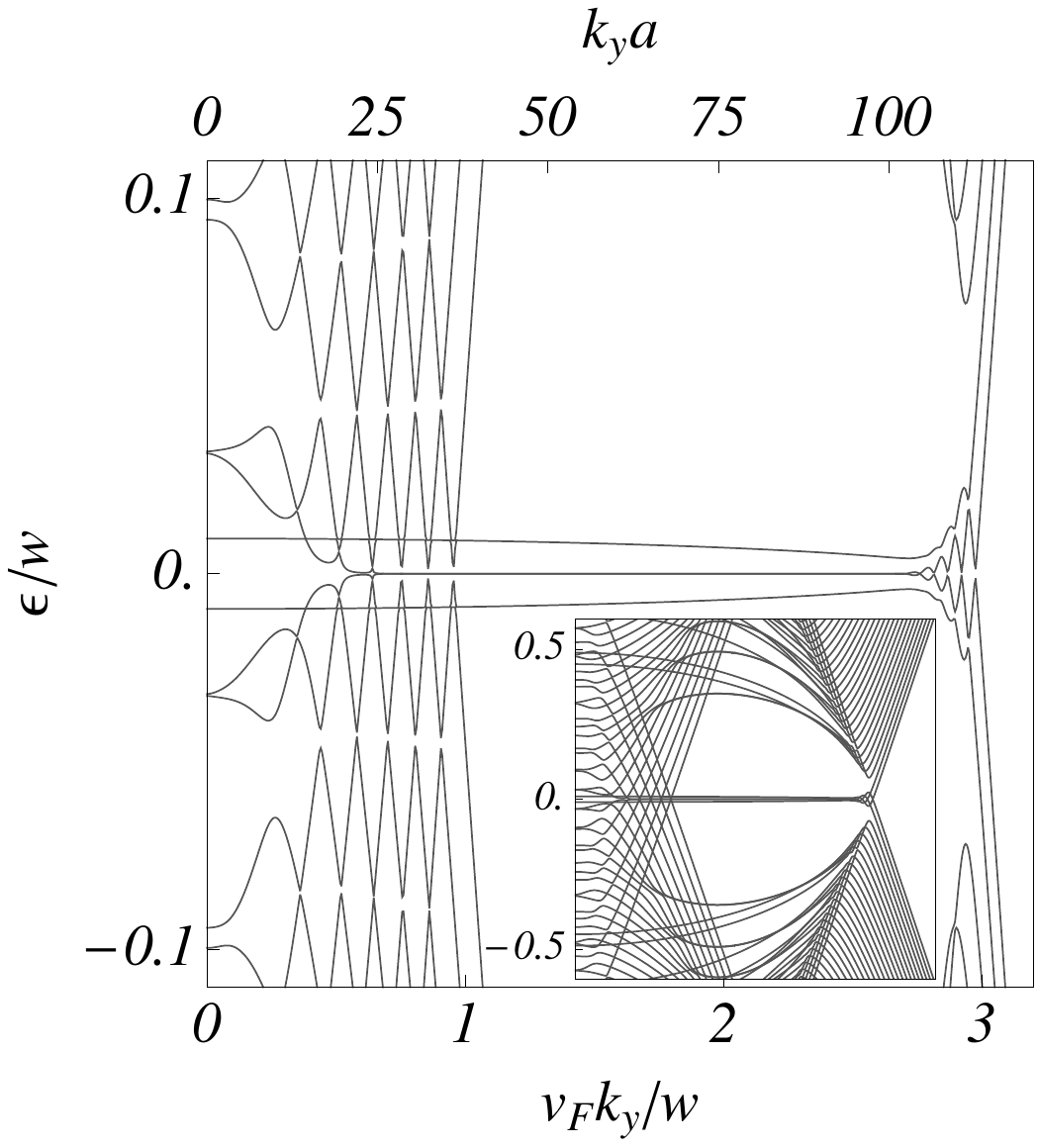}
\includegraphics[height=4.7cm]{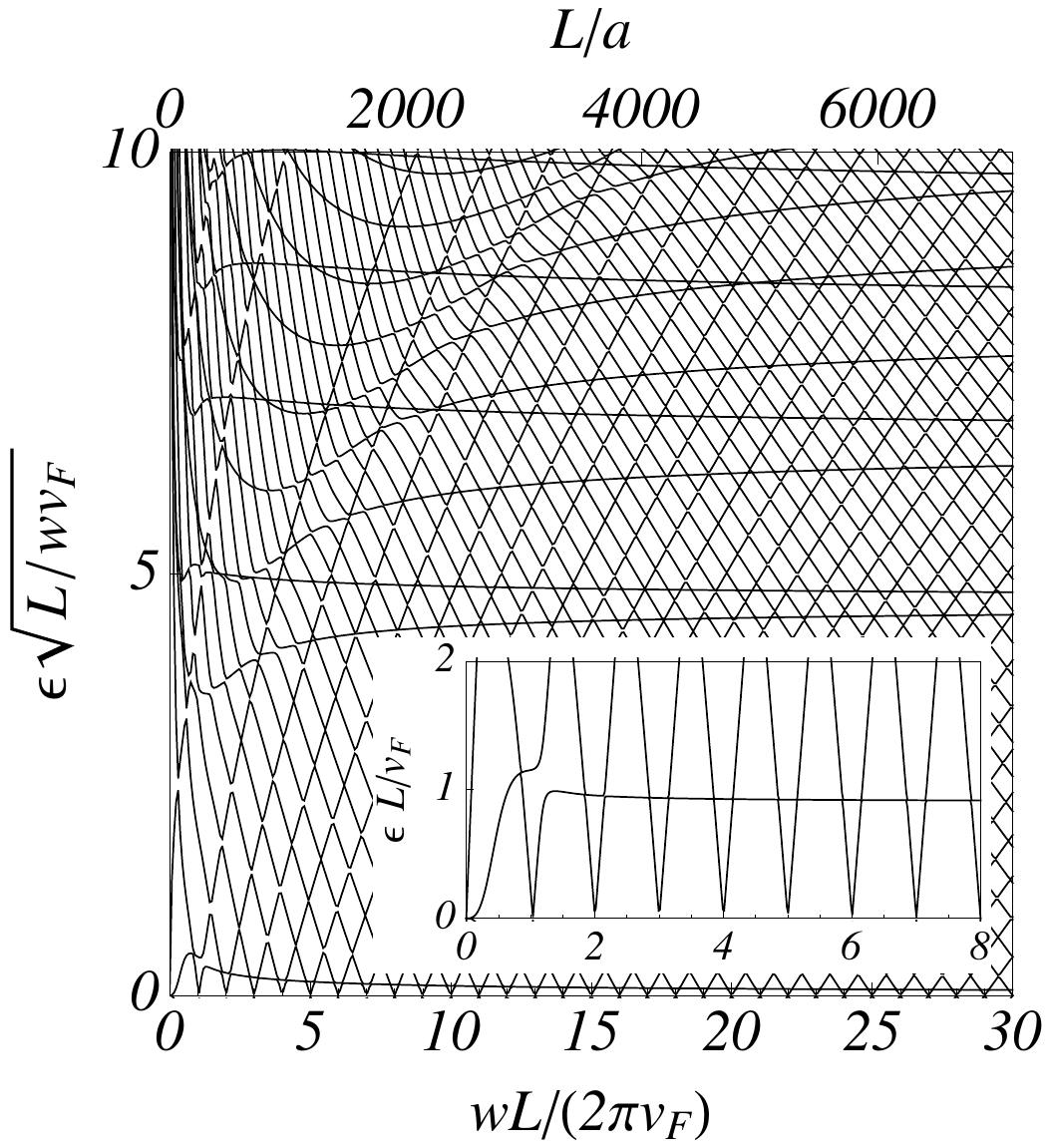}
\end{center}
\caption{Left:  Dispersion of the low-energy eigenstates of the Hamiltonian (\ref{HA}) as a function of $k_y$, for $k_x = 0$ and $L \sim 3700 a$, where $a$ is the C-C distance. Note the zero-energy band (confined between $AB'$ and $BA'$) and its satellite flat band (confined around $AA'$). The inset covers a larger energy range.
Right: Low-energy levels of the sheared bilayer as a function of the period
$L$ for $k_x = k_y = 0$ . Note the two types of states, $AA'$-confined, which scale as $\varepsilon\sim 1/\sqrt{L}$ (all but the first, which scales as $1/L$, see inset), and the $AB'-BA'$ states, which cross zero energy when $wL/v_F=2\pi n$ for integer $n$.}
\label{two}
\end{figure}

This electron system has the characteristic property of developing flat bands of spatially confined states at large $L$, whose formation is fully controlled by the effect of
the gauge potential $\hat{\mathbf{A}}$. Computing the energy levels of the Hamiltonian (\ref{HA}), one
observes that at large $L$ the system develops two increasingly narrow subbands around zero energy of states confined between $AB'$ and $BA'$ regions, see Fig. \ref{two}. Their energy, for any given momentum $k_x$ and $|k_y|\lesssim 3w/v_F$, oscillates towards zero, crossing it periodically as $L$ increases (e.g. whenever $w L/2\pi v_F$ is an integer if $k_y=0$, see inset on the right panel of Fig. \ref{two}).
Additionally, a second pair of flat bands appear at a finite energy, corresponding to states confined around $AA'$. All these bands become $AA'$-confined and linearly dispersive in $k_y$ for $|k_y|\gtrsim 3w/v_F$, although they remain non-dispersive in the $x$ direction. These features are strongly reminiscent of the Landau-level to snake-state transition in carbon nanotubes of large radius
in a real perpendicular magnetic field \cite{Perfetto:PRB07}, which have also an
effectively modulated magnetic flux.


This confinement phenomenology may be understood from the effect of a confining potential
created purely by the gauge field $\hat{\mathbf{A}}$.
The equation for the eigenstates $\Psi$ of $H$ can be expressed after squaring
the Hamiltonian (and disregarding for simplicity the scalar potential at this point) as
\begin{equation}
(-\boldsymbol{\partial}^2 + i \boldsymbol{\partial} \cdot \hat{\mathbf{A}} + 2i \hat{\mathbf{A}} \cdot \boldsymbol{\partial}
+ A_x^2 + A_y^2 - \sigma_z \hat{F}_{xy} ) \Psi  =  (\varepsilon /v_F)^2  \Psi
\label{D2}
\end{equation}
where the field strength is conventionally defined in terms of the matrix-valued
potential $\hat{A}_{\mu}$ as
$\hat{F}_{\mu \nu} = \partial_{\mu} \hat{A}_{\nu} - \partial_{\nu} \hat{A}_{\mu}  -  i [\hat{A}_{\mu},\hat{A}_{\nu}]$.
Given the invariance of $H$ under the combined operation of charge conjugation and parity,
the eigenstates can be chosen in the form
$\Psi(\mathbf{r})=\left[\phi_1^\star(-\mathbf{r}), \phi_1(\mathbf{r}), \phi_2^\star(-\mathbf{r}), \phi_2(\mathbf{r})\right]$, for some $ \phi_{1,2}$.
In the limit of zero transverse momentum $k_y$, the combinations
$\phi_\pm(\mathrm{r})\equiv\phi_1(\mathbf{r})\pm \phi_2^\star(-\mathbf{r})$ decouple, and
the above equation translates, at large $L$, into
\begin{equation}
-v_F^2\phi_\pm''(x) =  -V^\pm_{\mathrm{eff}}(x)\phi_\pm(x) + \mathcal{O}\left(\frac{v_F}{w L}\right)
\label{Veff}
\end{equation}
with $V^\pm_{\mathrm{eff}}(x)  \equiv  -(\pm \varepsilon+A_x+A_y)(\pm \varepsilon+A_x-A_y)$ \footnote{In Eq.
(\ref{Veff}), we have neglected corrections of the order $\partial_x A_j(x)\sim w/L$,
although not $-i\partial_x\phi^\star(-x)=(\varepsilon-V_{BA'}(x))\phi(x)$}.
This is the wave equation of a scalar mode with eigenvalue $E=0$ under the influence of an
$\varepsilon$-dependent confining potential $V^\pm_{\mathrm{eff}}(x)$, sketched in Fig. \ref{one}.
$\varepsilon=0$ eigenstates centered around $AB'$ and $BA'$ regions will arise  whenever
a level of such potential crosses $E=0$. Such states will be peaked exactly at $AB'$ and $BA'$,
since the well has $E=0$ turning points at said regions. Moreover, a discrete set of $E=0$
eigenstates centered around the $AA'$ local minimum will arise at energy $\varepsilon\sim 1/\sqrt{L}$.
These two types of states are apparent in the numerical bandstructure plotted
on the right panel of Fig. \ref{two}.

The above analysis in terms of $V_\mathrm{eff}$ relies crucially on the non-Abelian character of the gauge potential, $[\hat{A}_x,\hat{A}_y]\neq 0$. Without this property, the recurrence of zero-energy states as $L$ increases would not appear. This may be appreciated from an alternative point of view. In order for a (normalizable) zero-energy state to exist, the operator $W_{\varepsilon=0}$ relating the wavefunction at $x=0$ and $x=L$, $[\phi_1(L),\phi_2(L)]=W_{\varepsilon=0}[\phi_1(0),\phi_2(0)]$, must have at least one eigenvalue of modulus one. Since at zero energy Eq. (\ref{HA}) leads to
\[
-i\partial_x\left(\begin{array}{cc}\phi_1\\\phi_2\end{array}\right)=(ik_y+A_x\tau_1-iA_y\tau_2)\left(
\begin{array}{cc}\phi_1\\\phi_2\end{array}\right),
\]
we have for $k_y=0$
\[
W_{\varepsilon=0}={\rm Pexp}\left\{i\int^L_0 dx \: [A_x(x) \tau_1-i A_y(x) \tau_2]\right\}
\]
where  ``Pexp'' denotes the path-ordered product of exponentials of differential
line elements.\footnote{Incidentally, the transfer operator $W$ takes the form of an open Wilson loop, due to the first order character of the Dirac equation.} One can check that this operator becomes unitary when $wL/v_F = 2\pi n$ for integer $n$.
This is the condition for the existence of normalizable zero-energy modes, in agreement with the numerical results.

{\em Low energy description of twisted bilayers.---}
At energies below $1$ eV, a twisted bilayer may be accurately modeled by
Hamiltonian (\ref{H}), where the shift $\Delta K$ in the relative
position of the Dirac points in each layer comes as a consequence of the rotation
by the twist angle $\theta$. If we take the original position of the $K$
points as $\mathbf{K} = (4\pi /3a_0, 0)$, the shift in each layer
is given by $\pm \Delta \mathbf{K}/2 = (0, \pm K \sin(\theta/2))$.
On the other hand, $\theta$ also fixes the size of the Moir\'e pattern unit cell,
which grows as $\theta$ decreases. More precisely, the  Bravais superlattice
formed by the Moir\'e pattern has primitive vectors
$\mathbf{L}_{\pm}= L (\sqrt{3}/2,\pm 1/2) $, where
$L=a_0/2\sin(\theta/2)$. This periodicity becomes exact on an atomic level when
the rotation is \emph{commensurate} and \emph{minimal}, such that
$L=\sqrt{1+3n+3n^2}a_0$ for some integer $n>0$ \cite{LopesdosSantos:PRL07}.



The interlayer coupling may be written in terms of a single periodic profile
$V (\mathbf{r}) = V (\mathbf{r}+\mathbf{L}_+) = V (\mathbf{r}+\mathbf{L}_-)$,
in such a way that if we fix $V_{AA'} (\mathbf{r}) = V (\mathbf{r})$, then
$V_{AB'} (\mathbf{r}) = V (\mathbf{r} + (\mathbf{L}_+ + \mathbf{L}_-)/3)$ and
$V_{BA'} (\mathbf{r}) = V (\mathbf{r} - (\mathbf{L}_+ + \mathbf{L}_-)/3)$. A
common procedure is to assume that the interlayer hopping is dominated by
processes with momentum-transfer $\mathbf{Q}_0 = 0$ or equal to the reciprocal
vectors $\mathbf{Q}_{1,2} = (\pm 2\pi/\sqrt{3},2\pi)/L$ \cite{LopesdosSantos:PRL07,Bistritzer:PNAS11},
so that $V (\mathbf{r}) = (w/v_F) \sum_j \exp (i \mathbf{Q}_j \cdot \mathbf{r} )$. 
Coupling $V$ is complex in this case, however,  but we can still carry out the procedure
of the preceding section by defining $A_{1x} = -{\rm Re}(V_{AB'}+V_{BA'})/2,
A_{2x} = {\rm Im}(V_{AB'}+V_{BA'})/2, A_{1y} = {\rm Im}(V_{AB'}-V_{BA'})/2$ and
$A_{2y} = {\rm Re}(V_{AB'}-V_{BA'})/2$. We can then write the Hamiltonian for
the twisted bilayer as
\begin{equation}
H=v_F \boldsymbol{\sigma}\cdot(-i\boldsymbol{\partial}-\tau_3 \Delta \mathbf{K}/2 -\hat{\mathbf{A}})+v_F \hat{\Phi}
\label{DA}
\end{equation}
with non-Abelian potentials
$\hat{\mathbf{A}}=(A_{1x} \tau_1 + A_{2x} \tau_2 , A_{1y} \tau_1 + A_{2y} \tau_2 )$
and $\hat{\Phi} = {\rm Re} (V_{AA'}) \tau_1 - {\rm Im} (V_{AA'}) \tau_2$ \footnote{This procedure shows that
it is always possible to trade the complex couplings $V_{AA'}, V_{AB'}, V_{BA'}$ by the potentials
$\hat{\mathbf{A}}, \hat{\Phi}$, which encode 6 independent real functions taking into account the different
$\tau_1, \tau_2$ projections.}.

The mismatch $\Delta \mathbf{K}$ of the Fermi points may be removed by
carrying out a gauge transformation on the spinors,
$\Psi = \exp ( (i/2) \tau_3 \: \Delta \mathbf{K} \cdot \mathbf{r}) \widetilde{\Psi}$,
at the expense of introducing new potentials
$\widetilde{V}_{ij} ( \mathbf{r} )= V_{ij} ( \mathbf{r} ) e^{-i \Delta \mathbf{K}\cdot \mathbf{r}}$.
We finally get a modified expansion
$\widetilde{V}(\mathbf{r}) = (w/v_F) \sum_j \exp (i \mathbf{q}_j \cdot \mathbf{r} )$,
with a star of three vectors $\mathbf{q}_j$.  Note that $|\widetilde{V}(\mathbf{r})|=|V(\mathbf{r})|$ has conical singularities at the center of $AB'/BA'$ regions.

\begin{figure}[t]
\includegraphics[height=5.3cm]{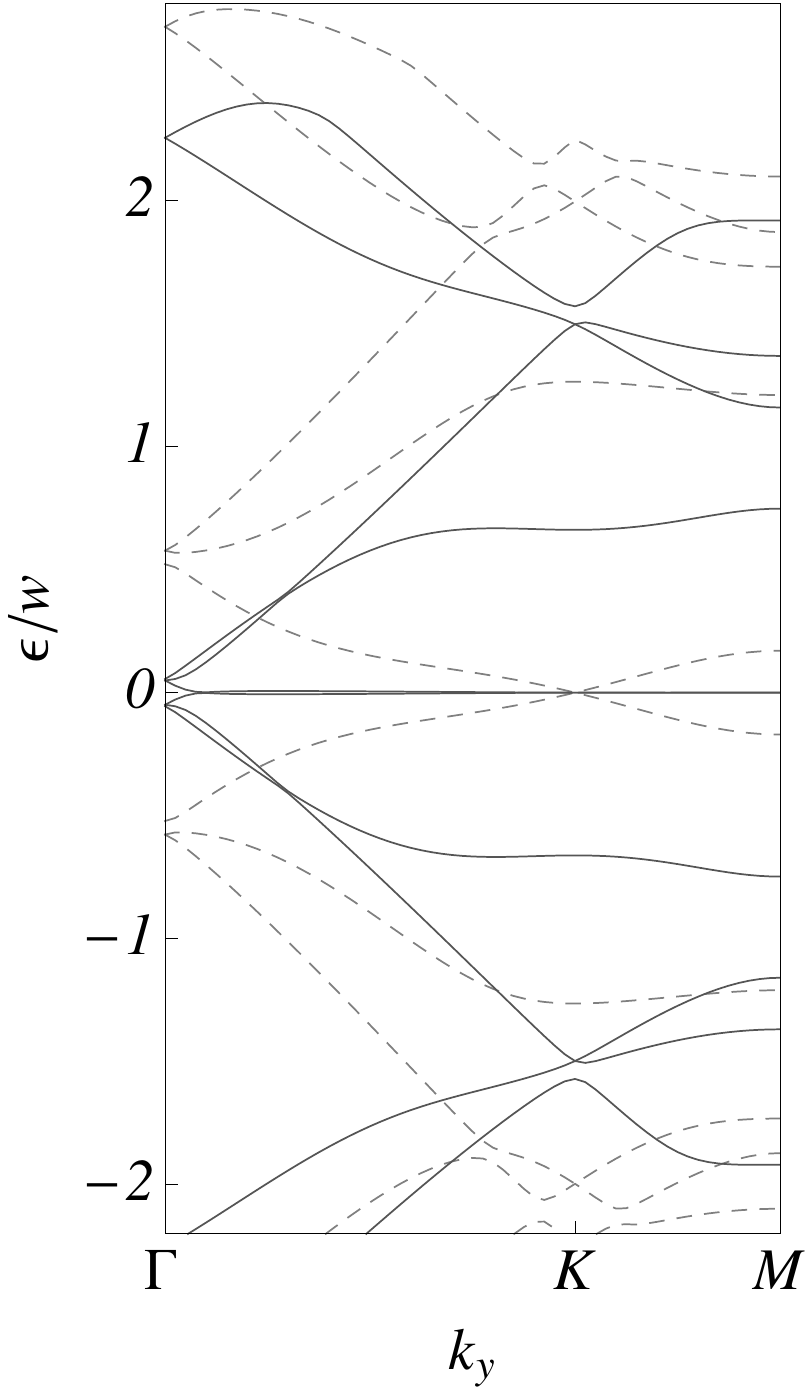}
\includegraphics[height=5.3cm]{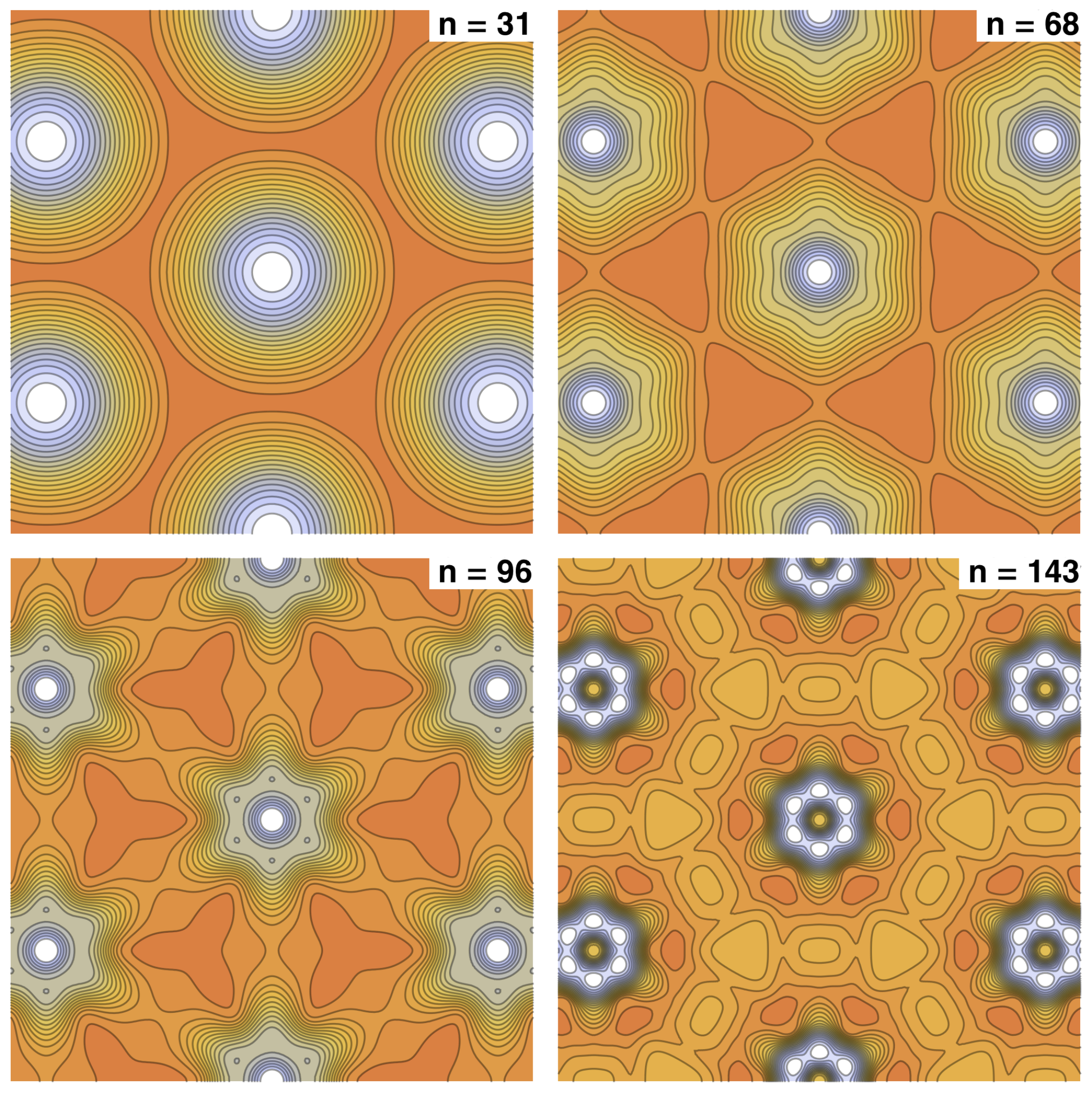}
\caption{Left: Low-energy subbands of the Hamiltonian (\ref{DA}) along the
first Brillouin zone of the bilayer superlattice for $n = 20$ (dashed lines) and $n = 31$
(full lines), for which a zero-energy band develops and the Fermi velocity at the $K$ point vanishes. Right: Localization pattern (in logarithmic color scale, white is maximum) around $AA'$ stacking of wavefunctions on the zero-energy band for the first four values of $n$ at which the Fermi velocity vanishes.}
\label{three}
\end{figure}

Two representative bandstructures obtained numerically from the Hamiltonian (\ref{DA})
for different values of $\theta$ are plotted in Fig. \ref{three}.
The first corresponds to an index $n = 20$ and exhibits a lowest subband
with vanishing energy at the two Dirac points originating from the graphene
layers. As the angle $\theta$ is decreased,
the energy scale of the lowest subband is significantly lowered, until
it becomes remarkably flat for values of $n$ around $n = 31$ ($\theta\approx 1^{\circ}$), exhibiting zero Fermi velocity at the $K$ point and a bandwidth that is more than 100 times smaller than the scale of the
next subband. (Note however that this is not a topological zero mode in the sense of a standard zero Landau level \cite{Katsnelson:PRB08}, since the Atiyah-Singer index \cite{Atiyah:TAOM68} is zero).
Lowering $\theta$ further, the lowest subband becomes dispersive
once more, before collapsing again, and so on, showing a recurrent behavior
as a function of the size $L$ of the Moir\'e pattern \cite{Bistritzer:PNAS11}.

For low values of $\theta$ ($n \gtrsim 31$), the lowest-energy
eigenstates show a strong confinement in the regions with $AA'$ stacking, as shown in Fig. \ref{three}, which is confirmed by atomistic tight binding calculations \cite{Trambly-de-Laissardiere:NL10}.
This confinement is essentially controlled by the vector potential $\hat{\mathbf{A}}$,
as the pattern of confinement remains unmodified when the scalar potential
$\hat{\Phi}$ is ideally switched off in the model. The eigenstates obey now
an equation similar to (\ref{D2}), but with $A_x^2 + A_y^2$ replaced by
$A_{1x}^2 + A_{2x}^2 + A_{1y}^2 + A_{2y}^2$ and Zeeman coupling to
$\hat{F}_{xy} = \partial_x A_{1y} \tau_1 + \partial_x A_{2y} \tau_2
  - \partial_y A_{1x} \tau_1 - \partial_y A_{2x} \tau_2
   +  2 A_{1x} A_{2y} \tau_3  -  2 A_{2x} A_{1y} \tau_3 $.
The contributions to the energy square of order $\sim w^2$ can be combined in
the form  $(A_{1x} \pm A_{2y})^2 + (A_{2x} \mp A_{1y})^2$. This function
becomes zero only at the center of $AA'$ stacking and at the center of either $AB'$
or $BA'$ stacking (depending on the eigenvalues of $\sigma_z$ and $\tau_3$).
This degeneracy is broken by the derivative terms in $\hat{F}_{xy}$, which tend to confine at
points where the gradients of $\widetilde{V}_{AB'}$ and $\widetilde{V}_{BA'}$
become higher. These functions become flatter at the
regions of $AB'$ and $BA'$ stacking, respectively, and are more steep
at the center of $AA'$ stacking, explaining the effect exerted by the
vector potential to confine in the latter region.



We note that the first instance at which the lowest subband
becomes flat has a simple interpretation as the situation where the analogue of the
magnetic length $l_B \sim \sqrt{v_F L/w}$ starts to fit in the bilayer supercell of size $L$.
One can actually check that, at $n = 31$, the result of
computing the flux integral $\hat{\varphi} = \int d^2 r \hat{F}_{xy} $
leads to values
$\hat{\varphi}\approx \Phi_0 \tau_2, \Phi_0(\cos (\pi/6) \: \tau_1 - \sin (\pi/6) \: \tau_2)$,
and $-\Phi_0 (\cos (\pi/6) \: \tau_1 + \sin (\pi/6) \: \tau_2)$ for supercells rotated by
$2\pi/3$ in the twisted bilayer, with $\Phi_0=2\pi$ (in units $\hbar=1$). This corresponds to the flux quantum rotated in the
SU(2) flavor space.
Unlike for that first instance, higher values of $n$ giving rise to a flat lowest-energy subband do depend on the strength of the $V_{AA'}$ coupling \footnote{To be precise, while a finite value of $V_{AA'}$ does not destroy subsequent instances of flat-band formation, it does lower their corresponding $L$ as compared to the case $V_{AA'}=0$, for which they satisfy the simple relation $wL/v_F\approx 2\pi\left(j+\frac{1}{2}\right)$, for integer $j$. Moreover, a finite $V_{AA'}$ renders the lowest subband with a small residual bandwidth that is non-existent in the pure magnetic case.}.
However, the essential spatial confinement properties of the corresponding lowest-energy eigenstates do not. They remain confined around $AA'$ stacking. They also acquire higher angular momentum components and become increasingly ring-shaped for higher values of $n$ (see Fig. \ref{three}), as expected for the excited states of a 2D potential well centered around $AA'$ stacking.


Experimental measures of the low-energy electronic properties of twisted
bilayers have been reported in particular in Ref. \onlinecite{Luican:PRL11}. It has been
found that, at a certain value $\theta \sim 1^{\circ}$, the renormalized Fermi velocity
near the $K$ point of the twisted bilayer
becomes so small that the picture based on Dirac quasiparticles breaks down.
This comes together with the observation of a clear
pattern of spatial confinement in the local density of states, which adopts the
form of a triangular charge density wave following the modulation of the
Moir\'e pattern. These features are fully consistent with the confinement
of the low-energy eigenstates in the regions of
$AA'$ stacking due to the action of the gauge potential, which provides a strong confinement mechanism according to the preceding
discussion. This single-particle mechanism will cooperate with the additional many-body effects that may also contribute to the modulation of the charge in the system.

{\em Conclusion.---}
We have shown that the Moir\'e-like modulation of the interlayer hopping in graphene
bilayers leads to a very rich phenomenology, which can be described in terms
of effective non-Abelian gauge potentials in the low-energy electronic theory. We have shown that any additional terms arising from the stacking modulation, such as non-Abelian scalar potentials, do not qualitatively modify the low energy electronic structure.
In the case of sheared bilayers with quasi-1D Moir\'e patterns,
the gauge potential is equivalent to a confining potential that leads to low-energy charge accumulation along 1D strips. The effect of the non-Abelian gauge potential in rotationally faulted bilayers is also to develop a characteristic spatial pattern of confinement, and the formation of dispersionless bands for discrete value of the Moiré periods. We conclude these two effects are the characteristic signature of Moiré-induced non-Abelian gauge potentials in graphene bilayers.

The emergence of these types of gauge fields is generic to systems of coupled Dirac equations, and the analysis presented here can be extended to multilayered systems with SU(N) gauge groups. One may also furthermore envision the possibility of tuning the non-Abelian fields caused by stacking by applying generic strain fields to Moiré bilayers. These will not only give rise to Abelian fields as in monolayers, but also to small modifications of the stacking non-Abelian fields \cite{Mariani:11}, whose interplay is known to produce a rich phenomenology \cite{Estienne:NJOP11}.

We acknowledge financial support from MICINN (Spain) through grants
FIS2008-00124 and CONSOLIDER CSD2007-00010. This research was supported in part by the National Science Foundation under Grant No. PHY11-25915.

\bibliography{biblio}

\end{document}